\documentclass[%
reprint,
showpacs,preprintnumbers,
 amsmath,amssymb,
 aps,
 prd,
]{revtex4-1}

\usepackage{graphicx}
\usepackage{epstopdf}
\usepackage{makebox}

\begin{document}


\title{Parameter Limits for Neutrino Oscillation with Decoherence in KamLAND}


\author{G. Balieiro Gomes}
\email[]{balieiro@ifi.unicamp.br}
\affiliation{Instituto  de  F\'isica  Gleb  Wataghin\\  Universidade  Estadual  de  Campinas - UNICAMP\\ Rua S\'ergio Buarque de Holanda, 777 \\  13083-970,  Campinas,  S\~ao Paulo,  Brazil}

\author{M. M. Guzzo}
\email{guzzo@ifi.unicamp.br}
\affiliation{Instituto  de  F\'isica  Gleb  Wataghin\\  Universidade  Estadual  de  Campinas - UNICAMP\\ Rua S\'ergio Buarque de Holanda, 777 \\  13083-970,  Campinas,  S\~ao Paulo,  Brazil}

\author{P. C. de Holanda}
\email{holanda@ifi.unicamp.br}
\affiliation{Instituto  de  F\'isica  Gleb  Wataghin\\  Universidade  Estadual  de  Campinas - UNICAMP\\ Rua S\'ergio Buarque de Holanda, 777 \\  13083-970,  Campinas,  S\~ao Paulo,  Brazil}

\author{R. L. N. Oliveira}
\email{robertol@ifi.unicamp.br}
\affiliation{Instituto  de  F\'isica  Gleb  Wataghin\\  Universidade  Estadual  de  Campinas - UNICAMP\\ Rua S\'ergio Buarque de Holanda, 777 \\  13083-970,  Campinas,  S\~ao Paulo,  Brazil}


\date{\today}

\begin{abstract}
In the framework of Open Quantum Systems we analyze data from KamLAND
by using a model that considers neutrino oscillation in a three-family
approximation with the inclusion of the decoherence effect. Using a
$\chi^2$ test we find new limits for the decoherence parameter which
we call $\gamma$, considering the most recent data by KamLAND.
Assuming an energy dependence of the type $ \gamma = \gamma_0 \left(
E/E_0 \right) ^n$, in 95 \% C.L. the limits found are $3.7 \times
10^{-24} GeV$ for $ n=-1$, $6.8 \times 10^{-22} GeV$ for $ n=0$, and
$1.5 \times 10^{-19} GeV$ for $ n=1 $ on the energy dependence.
\end{abstract}


\maketitle


\section{Introduction}

In general, the study of vacuum neutrino oscillations is made in the
framework of usual Quantum Mechanics, which considers the neutrino
system as isolated. In this work we will do a different kind of
analysis, in the framework of Open Quantum Systems, considering that
the neutrinos, which will be our subsystem of interest, have a
coupling with the enviroment.

The theory of Open Quantum Systems was created to deal with the case
in which the system of interest is not considered isolated
\cite{breuer, 28, ellis}. Instead, it has a coupling with the enviroment,
and such coupling has important consequences on its evolution.

As we will see, the coupling with the enviroment will act changing the
superposition of states, eliminating the coherence, similarly to what
we have when a measurement is made in a quantum system, and generating 
a decoherence efffect. We can find in the
literature studies of the decoherence effect applied to neutrino
oscillations \cite{roberto, roberto2, roberto3, lisi}.

Using this different approach to study neutrino oscillations we see
that different forms of the survival probability are obtained
\cite{roberto}. The goal of this work is to test one of these forms 
using data from the KamLAND experiment.

KamLAND \cite{kl1, kl2, klprecision, klnew, klor} is a Long Baseline
experiment, located at the Kamioka mine, Gifu, Japan, and detects
electron antineutrinos which come from nuclear reactors being at an
average distance of $\sim180$km from the detector. It was constructed
to test the so called Large Mixing Angle (LMA) solution to the solar
neutrino problem, and its results were found to have a striking
agreement with solar neutrino results \cite{klor}.

The goal of this work is to obtain new limits for the parameter
$\gamma$ which describes decoherence, considering the most recent
KamLAND data. We will also stress its relevance and the difference
between the results found in this work from others such as the one
from \cite{lisi}. In Section I we review the Theory of Open Quantum
Systems, and we show how it can be used to study neutrino
oscillations. We present how the decoherence
effect arises, generating a different form of the survival
probability, which is tested using a $\chi^2$ test. 
The simulation results and the limits of the parameters are presented 
in Section II. We present our conclusions in Section III. 

\section{Formalism}

In this section we will introduce the formalism used to obtain
probabilities with dissipation effects from the Lindblad Master
Equation. In this formalism the neutrinos are treated as an open
quantum system and it interacts with the quantum environment. We
assume that the quantum environment works as a reservoir. These two
quantum states compose the global system, and from the interaction
between neutrinos and environment arise the dissipation effects
\cite{28, breuer}. In Open Quantum System theory it is possible to
show that if the interaction between the subsystem of interest, which
are the neutrinos in this case, and the reservoir is weak, the dynamic
can be obtained by the Lindblad Master Equation \cite{28, breuer}. A
reviewof the fundamentals of quantum open system theory can be found
in the following Refs. \cite{28, breuer}.

The Lindblad Master Equation can be written as \cite{12, 13}:
\begin{equation}
\frac{d}{dt} \rho (t) =	 -i [H,\rho(t)] + D[\rho(t)]
\label{i}
\end{equation}
with,
\begin{equation}
D[\rho(t)] =  \frac{1}{2} \displaystyle \sum_{k=1}^{N^2-1} \left( [ V_k , \rho(t) V_k^\dag ] + [ V_k \rho(t) , V_k^\dag ] \right) \,,
\label{ii}
\end{equation}
where $N$ is the dimension of the Hilbert space of the subsystem of
interest and $V_{k}$ describes the interaction between the subsystem
of interest and the environment. In this equation we see a term which
is equal to the one we have in the Liouville Equation, but we also
have the term $D[\rho(t)]$ which appears because we are dealing with
an open system, different from what we have in usual Quantum
Mechanics, where the system is considered isolated. 
$D[\rho(t)]$ must satisfy some mathematical constraint and then, it
can be phenomenologically parameterized.

We will impose on this equation that the entropy increases with time,
in order that $D[\rho(t)]$ evolves a pure state asymptotically to a
state of maximal mixing. Using the Von Neumann entropy it is possible
to show that this condition leads to restrictions on the operator
$V_{k}$, in particular we see that it must be Hermitian \cite{22}.

The Lindblad Equation in (\ref{i}) can be expanded in the basis of
$SU(3)$ matrices, since the three neutrino families are considered in
this work. In this form, each operator in Eq. (\ref{i}) can be
expanded as $O=a_{\mu}\lambda_{\mu}$, where $\lambda$ are the
Gell-Mann matrices. Then, the evolution equation in Eq. (\ref{i}) can
be written as  
 \begin{equation}
\frac{d}{dx}\rho_{k}(x)=2\epsilon_{ijk}H_{i}\rho_{j}(x)+D_{k l}\rho_{l}(x)\,,
\label{iii}
\end{equation}
 and the probability conservation leads to $D_{\mu 0} =D_{0\nu}=0$.

It is important to note that the $\dot{\rho}_{0}(t)=0$ and its
solution is given by $\rho_{0}(t)=1/N$, where N is the number of
families. For simplicity, we do not include this component in the
equation above.

There are many parameters in the dissipator matrix $D_{k l}$. However,
it is possible to reduce the number of these parameters considerably
if we impose some physical and mathematical constraints.  

In order to obtain a dissipator matrix $D_{k l}$ with parameters that describe well known effects, we can impose first that $[H,V_{k}]=0$. From the physical point of view, this commutation relation implies energy conservation in the neutrino subsystem and also this constraint includes the decoherence effect in the evolution. This effect eliminates the quantum coherence, and the oscillation probability is changed by damping terms that are multiplied by oscillation terms. In this condition, the $D_{k l}$ assumes the following form
\begin{equation}
D_{kl}=-{\rm diag}\{\gamma_{21},\gamma_{21},0,\gamma_{31},\gamma_{31},\gamma_{32},\gamma_{32},0\} \,,
\label{iv}
\end{equation}
where each $\gamma_{ij}$ can describe the decoherence effect between
the families $i$ and $j$ \cite{roberto2}.

Once the neutrinos are free to interact with the reservoir the energy
in the neutrino sector can fluctuate, and hence the energy conservation
constraint may not be satisfied. We can relax this constraint
adding other two new parameters in $D$, $D_{33}$ and $D_{88}$, such that the
dissipator in Eq. (\ref{iv}) becomes
\begin{equation}
D_{kl}=-{\rm diag}\{\gamma_{21},\gamma_{21},\gamma_{33},\gamma_{31},\gamma_{31},\gamma_{32},\gamma_{32},\gamma_{88}\} 
\label{v}
\end{equation}
where again $\gamma_{ij}$ can describe the decoherence effect between
the families $i$ and $j$, while $\gamma _{33}$ and $\gamma_{88}$
describe the so called relaxation effect.

The relaxation effect is a phenomenon that dynamically leads the
states to their maximal mixing state. This phenomenon appears in the
oscillation probabilities through the damping term multiplied by terms
that depend only on mixing parameters. Then, when the relaxation
effect is taken into account the probabilities tend asymptotically to
$1/N$, where $N$ is the number of families initially considered.

In general, if a particular density matrix represents an initial
physical state, the density matrix evolved by Eq. (\ref{i}) may not be
a well-defined quantum state. Complete positivity is a constraint on $
D_{k l}$ which always keeps the evolution made by Eq. (\ref{i}) as
being physical \cite{ben,28}. From complete positivity the $D_{k l}$
needs to be a positive matrix and this is satisfied if the diagonal
elements of $ D_{k l}$ are larger than the off-diagonal elements. So, we
are going to consider the dissipator matrix obtained in Eq. (\ref{v})
to evolve the neutrinos according to the complete positivity, which
corresponds to the most effective dissipator that we can obtain. Any
other off-diagonal element can be represented in function of the main
diagonal elements, since the $\gamma _{33}$ and $\gamma_{88}$
parameters are non-null.

The Hamiltonian in the effective mass basis can be written as 
\begin{eqnarray}
\tilde{H} &=&\frac{1}{2E}
\left(\begin{array}{c c c } 
\tilde{m}^2_{1} & 0 & 0 \\
0 & \tilde{m}^2_{2} & 0\\
0 & 0 & \tilde{m}^2_{3}  \end{array} \right)\,,
\label{ham}
\end{eqnarray}
with 
\begin{eqnarray}
\tilde{m}_{1}&=&-\frac{1}{2}\left((\delta \cos 2 \theta_{12} - A\cos^{2}\theta_{13})^{2} + \delta^{2} \sin^{2} 2\theta_{12}\right)^{\frac{1}{2}}\,,\nonumber \\
\tilde{m}_{2}&=&\frac{1}{2}\left((\delta \cos 2 \theta_{12} - A\cos^{2}\theta_{13})^{2} + \delta^{2} \sin^{2} 2\theta_{12}\right)^{\frac{1}{2}}\,,
\label{vi}
\end{eqnarray}
where $\delta=m^{2}_{2}-m^{2}_{1}$, $A=2\sqrt{2}n_{e}E \cos^{2} \theta_{13}$ and
\begin{equation}
\tilde{m}_{3}=\frac{1}{2}(2m^{2}_{3}-m^{2}_{2}-m^{2}_{1}+A \sin^{2}\theta_{13})\,.
\label{vii}
\end{equation}

The relation between the flavor state and the effective mass basis is given by the following transformation:
\begin{equation}
\rho_{\alpha}=U \rho_{\tilde{m}} U^{\dag}= U_{13} \tilde{U}_{12}\rho_{\tilde{m}} \tilde{U}^{\dag}_{12} U^{\dag}_{13}, 
\label{viii}
\end{equation}
where the $\rho_{\alpha}$ is the flavor state and $\rho_{\tilde{m}}$ is the effective mass state. The mixing matrix $U$ is explicitly defined as
\begin{eqnarray}
U &=&
\left(\begin{array}{c c c } 
\cos \theta_{13} & 0 & \sin \theta_{13} \\
0 & 0 & 0\\
-\sin \theta_{13} & 0 & \cos \theta_{13} \end{array} \right)
\left(\begin{array}{c c c } 
\cos\tilde{\theta}_{12} & \sin\tilde{\theta}_{12} & 0 \\
-\sin\tilde{\theta}_{12} & \cos\tilde{\theta}_{12} & 0\\
0 & 0 & 1  \end{array} \right)\,,\nonumber \\ &&
\label{ix}
\end{eqnarray}
and the effective mixing angle has the usual form
\begin{equation}
\sin^{2}2\tilde{\theta}_{12}=\frac{\delta^{2} \sin^{2}2\theta}{(\delta \cos 2 \theta_{12} - A\cos^{2}\theta_{13})^{2} + \delta^{2} \sin^{2} 2\theta_{12}}\,.
\label{x}
\end{equation}
 	
We have defined a diagonal form to the Hamiltonian in Eq. (\ref{ham}). Hence, the dissipators in Eq. (\ref{iv}) and Eq. (\ref{v}) remain diagonal as well.

The evolved density matrix in the effective mass basis is given by: 
\begin{eqnarray}
\rho_{\tilde{m}}(x)&= &
\left(\begin{array}{c c c } 
\rho_{11}(x) & \rho_{12}(0)e^{-\widetilde{\Delta}^{*}_{12}x} & \rho_{13}(0)e^{-\widetilde{\Delta}^{*}_{13}x} \\
\rho_{21}(0)e^{-\widetilde{\Delta}_{12}x} &\rho_{22}(x) & \rho_{23}(0)e^{-\widetilde{\Delta}_{23}x} \\
\rho_{31}(0)e^{-\widetilde{\Delta}_{13}x} & \rho_{32}(0)e^{-\widetilde{\Delta}^{*}_{23}x} &\rho_{33}(x)  \end{array} \right)\,,\nonumber \\ &&
\label{xi}
\end{eqnarray}
where $\rho_{ij}(0)$ are elements of the initial state obtained from Eq. (\ref{ix}) and  $\widetilde{\Delta}_{ij}=\gamma_{ij} +i(\tilde{m}^2_{i}-\tilde{m}^2_{j}) /2E$. While the $\rho_{ii}$ elements are given by
\begin{eqnarray}
\rho_{11}(x)&=&\frac{1}{3}+\frac{1}{2}e^{-\gamma_{33} x}\cos 2\tilde{\theta}_{12}\cos^{2}\theta_{13} \nonumber \\ && -\frac{1}{12}e^{-\gamma_{88}x} (1-3\cos2\theta_{13}) \,,\nonumber \\
\rho_{22}(x)&=&\frac{1}{3}-\frac{1}{2}e^{-\gamma_{33} x}\cos 2\tilde{\theta}_{12}\cos^{2}\theta_{13} \nonumber \\ && -\frac{1}{12}e^{-\gamma_{88}x} (1-3\cos2\theta_{13}) \,,\nonumber \\
\rho_{33}(x)&=&\frac{1}{3}+\frac{1}{6}e^{-\gamma_{88} x} (1-3\cos2\theta_{13})\,.
\label{xii}
\end{eqnarray}

These damping terms in the diagonal elements describe the relaxation
effect through the parameters $\gamma_{33}$ and $\gamma_{88}$. Besides
that, they depend on the mixing parameters $\theta_{12}$ and
$\theta_{13}$ and the distance between the source and the detection
point. The main diagonal in state (\ref{xi}) can be interpreted as the
probabilities to find $\tilde{m}_{1}$, $\tilde{m}_{2}$ or
$\tilde{m}_{3}$ of the observable $H$ in Eq. (\ref{ham}). In usual
quantum mechanics, these elements do not change within an adiabatic
propagation. So, analysing the state in (\ref{xi}) we can see how the
relaxation effect act on the probabilities.

The state in (\ref{xi}) shows how the relaxation effect depends on the
propagation distance. Considering MSW solution for solar neutrinos,
which produce a specific relation between mass eigenstates in the
final neutrino flux, we expect that the relaxation effects are
strongly constrained. We will present this analysis somewhere else,
but the Sun-Earth distance is of the order of $10^{17}$ eV$^{-1}$ and
a rough limit for both relaxation parameters is $~10^{-18}$ eV, in
order to have $\exp[-\gamma_{ii} x]\sim 1$. Thus, the analysis of
reactor neutrinos can disregard the relaxation effect because the
larger baseline to this source is much smaller than Sun-Earth
distance.

The off-diagonal elements are known as coherence elements. In state
(\ref{xi}), these elements tend to zero
during the propagation due to the damping terms. 
This is the exact definition of the
decoherence effect. But, in the solar neutrino context, these elements
are averaged out, and any decoherence effect information is lost
if we consider a model-independent approach \cite{robertosun}. 
Besides, since 
$|\Delta m^2_{13}|\sim |\Delta m^2_{23}| \gg |\Delta m^2_{12}|$, 
experiments such as KamLAND, that are tuned to test $\Delta m^2_{12}$, 
are not sensible to the coherence elements $\rho_{i3}$. These elements 
dependends on $\widetilde{\Delta}_{i3}x$ with $i\neq 3$, which
oscillate very fast, and hence are avereged out. 

So, disregarding the fast-oscillating terms and
the relaxation effects, the state is given by:
\begin{eqnarray}
\rho_{\tilde{m}}(x)&= &
\left(\begin{array}{c c c } 
\rho_{11}(0) & \rho_{12}(0)e^{-\widetilde{\Delta}^{*}_{12}x} & 0 \\
\rho_{21}(0)e^{-\widetilde{\Delta}_{12}x} &\rho_{22}(0) & 0 \\
0 & 0 &\rho_{33}(0)  \end{array} \right)\,,\nonumber \\ &&
\label{xiii}
\end{eqnarray}
and using the Eq. (\ref{ix}) to write the state above in the flavor basis, the survival probability can be obtained by taking
\begin{equation}
P_{\nu_\alpha \rightarrow \nu_\alpha} = Tr[\rho_{\alpha}(0) \rho_{\alpha}(t)] 
\label{xiv}
\end{equation}
where the initial state for $\bar{\nu}_{e}$ is $\rho_{\alpha}(0)=diag\{1,0,0\}$. So, the survival probability is given by \cite{klnew}:
 \begin{equation}
 P_{\nu_\alpha \rightarrow \nu_\alpha}^{3 \nu} = \cos^4(\theta_{13}) \tilde{P}_{\nu_\alpha \rightarrow \nu_\alpha}^{2 \nu} + \sin^4 (\theta_{13})
\label{xvi}
 \end{equation}
where $\tilde{P}_{\nu_\alpha \rightarrow \nu_\alpha}^{2 \nu}$ is written
\begin{equation}
 \tilde{P}_{\nu_\alpha \rightarrow \nu_\alpha}^{2 \nu} =  1 -\frac{1}{2} \sin^2 (2 \tilde{ \theta}_{12}) \left[ 1 - e^{- \gamma x} \cos  \frac{(\tilde{m}^2_{1}-\tilde{m}^2_{2})}{2E} x\right]
\label{xvii}
 \end{equation}
that is the same probability obtained in two-neutrino approximation when the decoherence effect is taken into account \cite{robertosun}.

It is important to explain the difference between the analysis made in
this work and the one made in \cite{lisi}, where they use a different
set of data from KamLAND (older than the one considered here), but
also consider data from solar neutrinos.
 
The first difference is that we are dealing with three neutrino
families. Moreover, as shown in \cite{roberto} and mentioned before,
there are cases in which, besides the decoherence effect, other
effects arise from the coupling with the enviroment, such as the so
called relaxation effect \cite{roberto}. Since in our case, as
previously shown, decoherence is the only relevant effect in the
interaction with the enviroment, including solar neutrinos in the
analysis would not bring any new information regarding the decoherence
parameter. Solar neutrinos cannot be used to bound decoherence,
because the fast oscillating terms in $\tilde{\Delta}_{ij}x $ average
out all the coherence terms~\cite{robertosun,farzan}. Therefore, the
effect studied here is different from the one studied
in~\cite{lisi}. According to \cite{robertosun}, the limits found 
in~\cite{lisi} are combined limits on relaxation and decoherence effects in a 
model-dependent approach.  We use a model-independent approach in this
paper to analyze the KamLAND data.

\section{Results}

We used the set of data presented in~\cite{klnew}, where the data is presented 
in 20 energy bins. For this set of data, we tested the usual oscillation scenario, and found for the best fit point: $\chi_{min}^2 = 22.96$, $\Delta m_{12}^2 = 8.05 \times 10^{-5} eV^2$, $\tan^2(\theta_{12}) = 0.40$. We can see that $ \chi_{min}^2 $ is close to the number of degrees of freedom, indicating a good agreement with the experimental data.

We considered now the oscillation probability in  Eq. (\ref{xvii}) with the three family approximation (Eq. 21), and the three free parameters  $\Delta m_{12}^2$, $\tan^2 (\theta_{12})$ and $\gamma$, also considering a possible energetic dependence on $\gamma$:

\begin{equation}
\gamma = \gamma_0 \left(  \frac{E}{E_0} \right) ^n ~~~,
\end{equation}
with $E_0 = 1 GeV$, such as the one done by \cite{lisi}. We did this test for $ n= 0$, $ n= 1$ and  $ n= -1$. 
We also considered the best-fit value for $\theta_{13}$ given by \cite{pdg}, $\sin^2(2 \theta_{13}) = 9.3 \times 10^{-2}$ .

The best-fit results for these scenarios can be seen in Table I, for the three values of $n$, where again we see that the value of $\chi_{min}^2$ is close to the number of degrees of freedom.

\begin{table*} 

\begin{ruledtabular}
\begin{tabular}{cccc}

	\hfill & $n=0$ & $n=1$ & $n=-1$ \\ \hline
  $ \chi_{min}^2$ &  21.44 & 21.92 & 21.03 \\ \hline
  $ \Delta m^2 $ &  $ 8.05 \times 10^{-5} eV^2$ &  $ 8.05 \times 10^{-5} eV^2$ &  $ 8.05 \times 10^{-5} eV^2$ \\ \hline
  $ \tan^2(\theta) $ &  0.44 & 0.42 & 0.47 \\ \hline
  $ \gamma_0 $ & $2.37 \times 10^{-22} GeV$ & $4.14 \times 10^{-20} GeV$ & $1.17 \times 10^{-24} GeV$ \\ 
\end{tabular}
\caption{Best-Fit Results For Three Free Parameters}
\vspace{1cm}
\end{ruledtabular}
\end{table*}

We can also see that including the third parameter $\gamma$ slightly improves the fit in comparison with the scenario where  $\gamma = 0$, with a decrease in the value of $\chi_{min}^2$.

We present confidence level curves for $n=0$ in the energy dependence, which can be seen in Figs. 1, 2 and 3, and in accordance to \cite{pdg} we chose the values of $ \Delta \chi^2 $ to get confidence levels of 68.27\%, 90\%, 95\%, 99\%, 99.73\% C.L.

\begin{figure}[tb]
\centering
\includegraphics[width=.45\textwidth,]{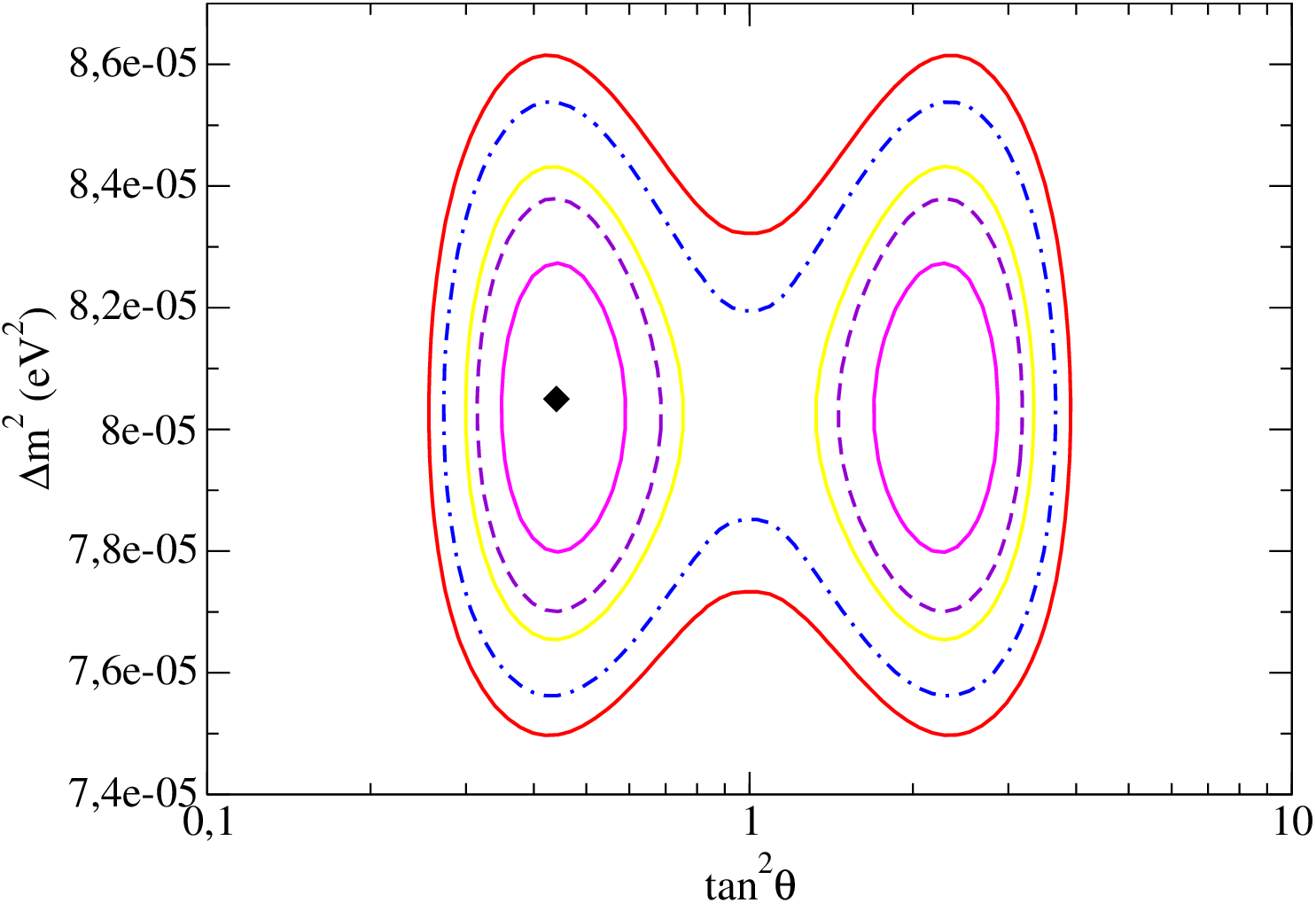}
\caption{Confidence Level curves for $n=0$. The curves correspond to 68.27\%, 90\%, 95\%, 99\% and 99.73\% C.L.}
\end{figure}

\begin{figure}[htb]
\centering
\includegraphics[width=.45\textwidth,]{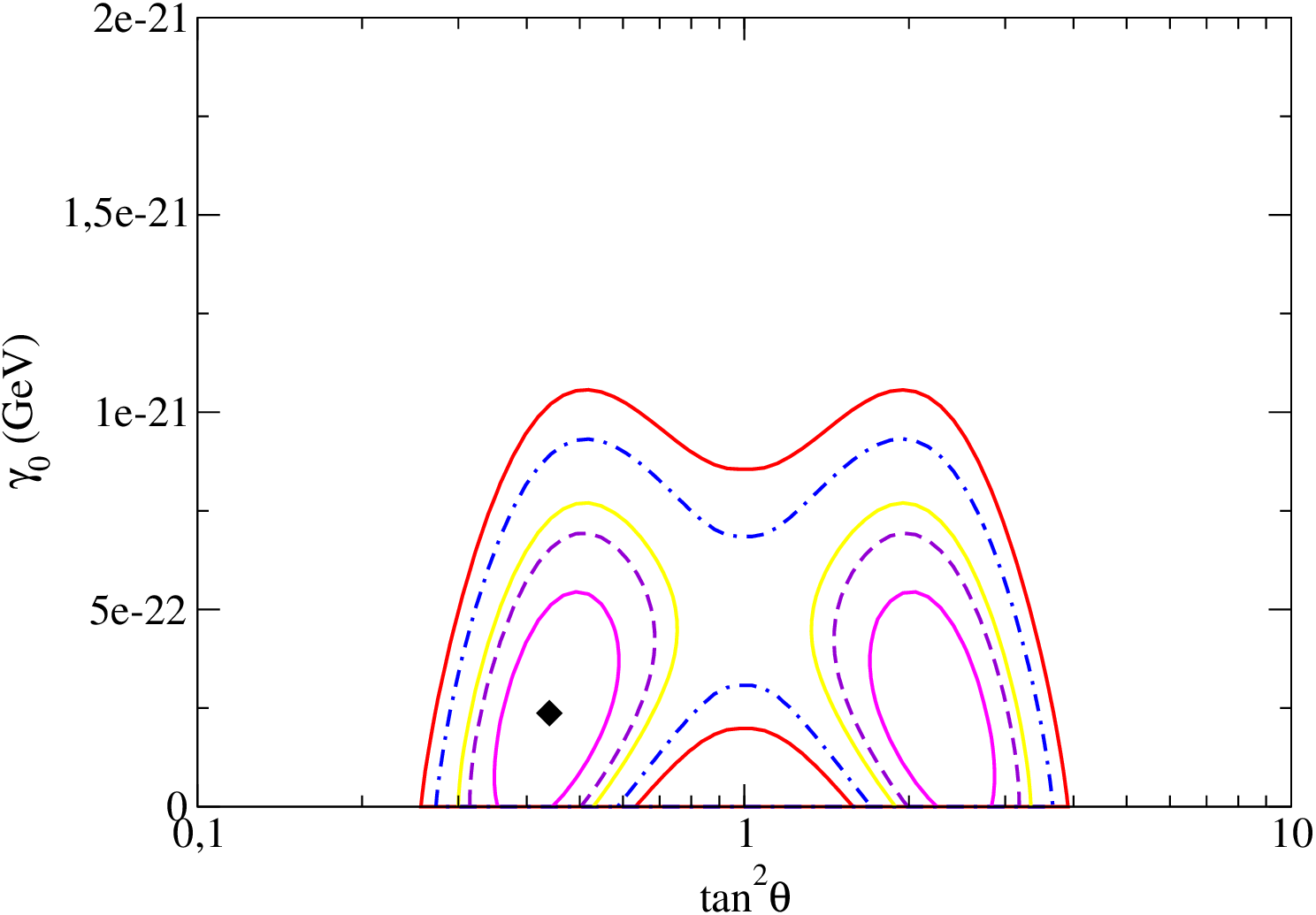}
\caption{Confidence Level curves for $n=0$. The curves correspond to 68.27\%, 90\%, 95\%, 99\% and 99.73\% C.L.}
\end{figure}

\begin{figure}[htb]
\centering
\includegraphics[width=.45\textwidth,]{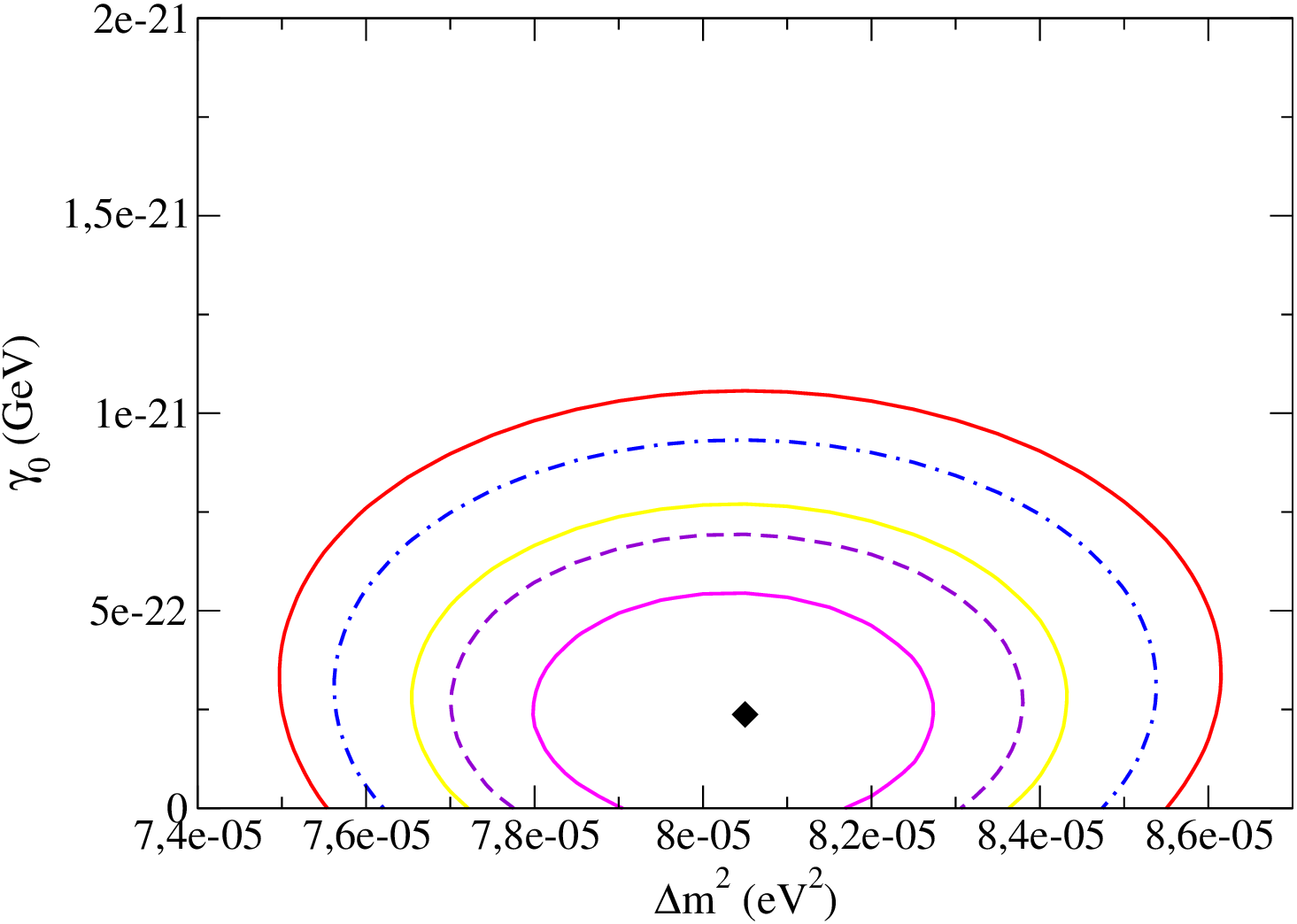}
\caption{Confidence Level curves for $n=0$. The curves correspond to 68.27\%, 90\%, 95\%, 99\% and 99.73\% C.L.}
\end{figure}

For $n=1$ in the energy dependence, the confidence level curves obtained are given in Figs. 4, 5 and 6, and for $n=-1$ in the energy dependence, the confidence level curves obtained are given in Figs. 7, 8 and 9:

\begin{figure}[htb]
\centering
\includegraphics[width=.45\textwidth,]{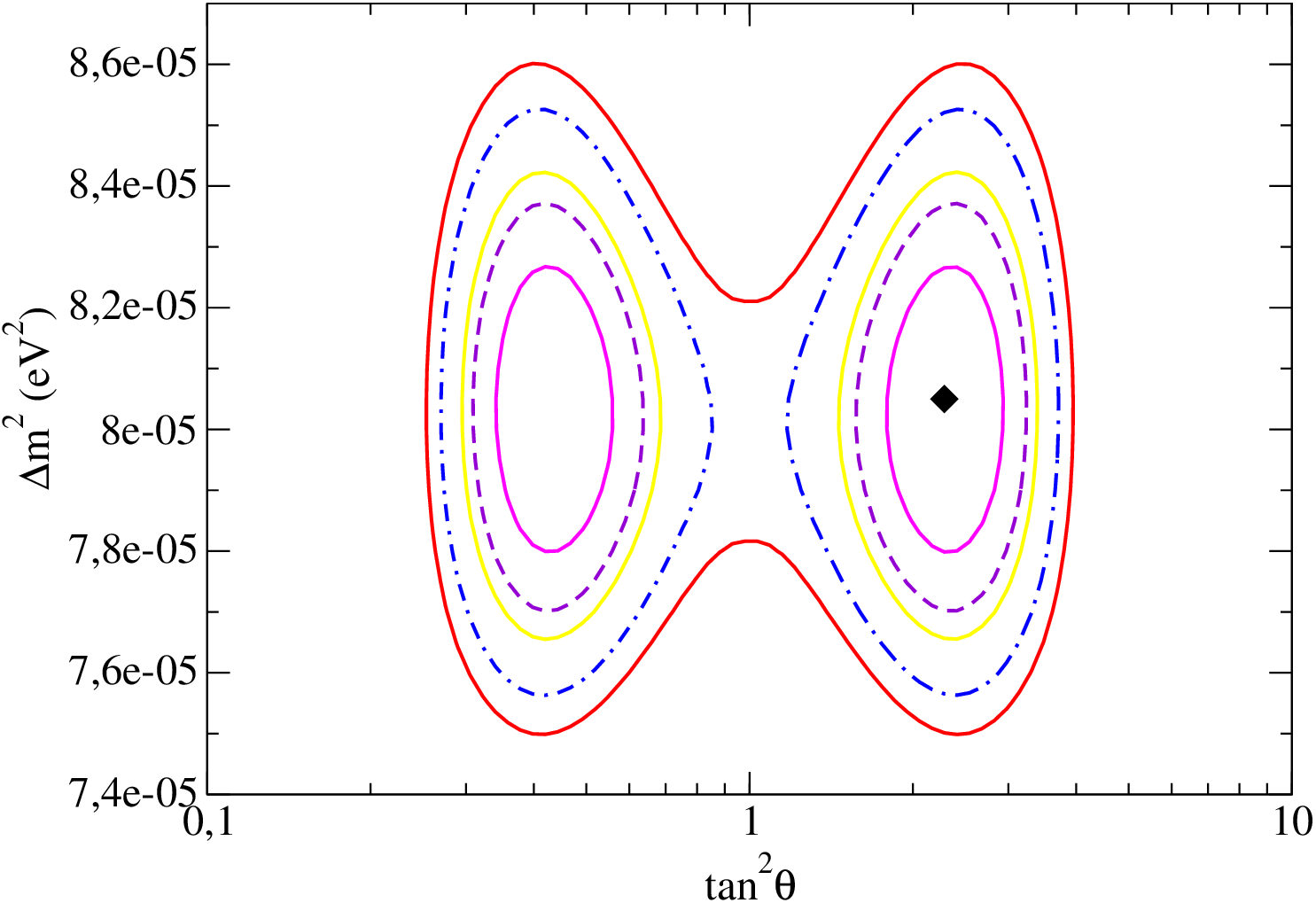}
\caption{Confidence Level curves for $n=1$. The curves correspond to 68.27\%, 90\%, 95\%, 99\% and 99.73\% C.L.}
\end{figure}

\begin{figure}[htb]
\centering
\includegraphics[width=.45\textwidth,]{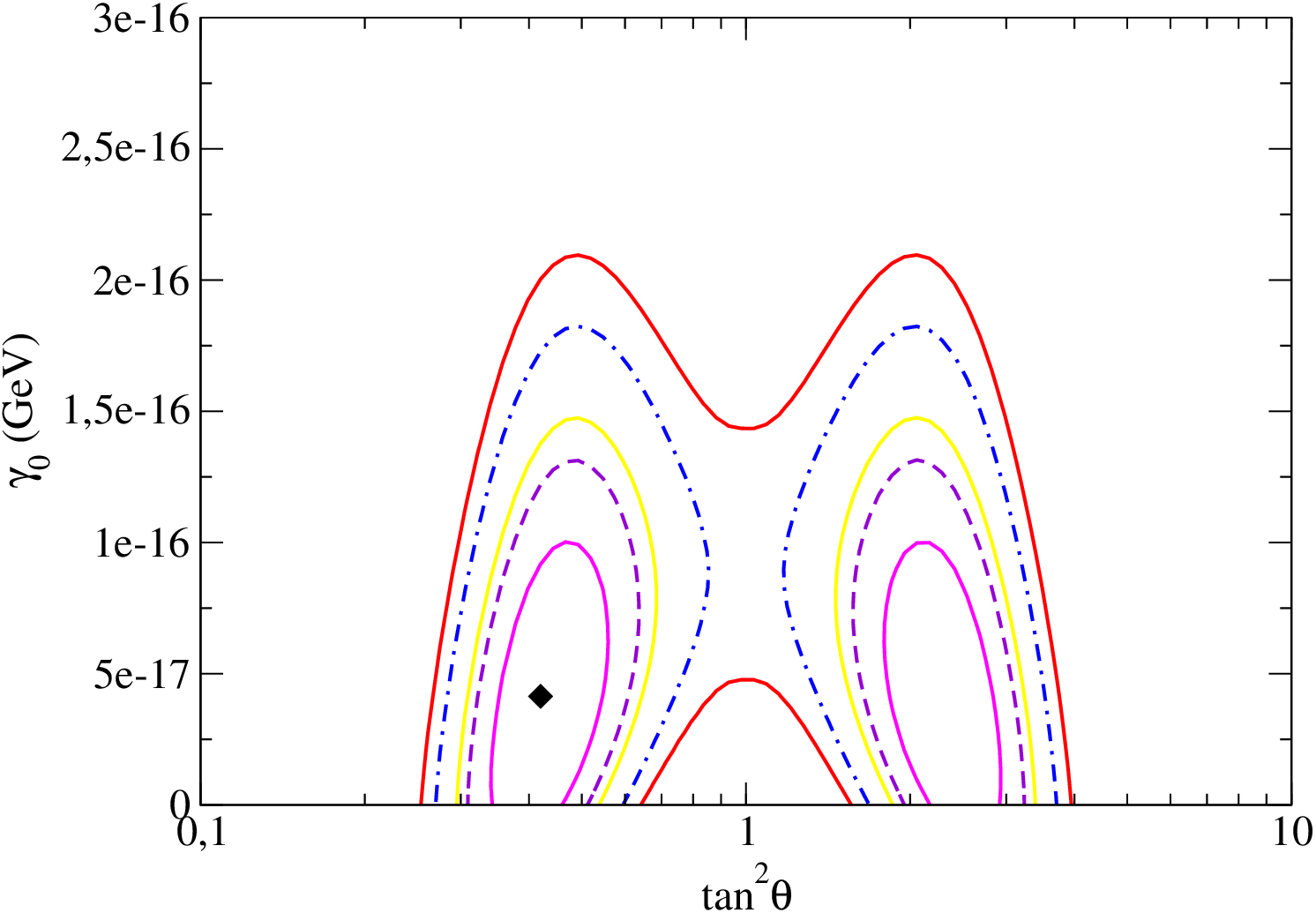}
\caption{Confidence Level curves for $n=1$. The curves correspond to 68.27\%, 90\%, 95\%, 99\% and 99.73\% C.L.}
\end{figure}

\begin{figure}[htb]
\centering
\includegraphics[width=.45\textwidth,]{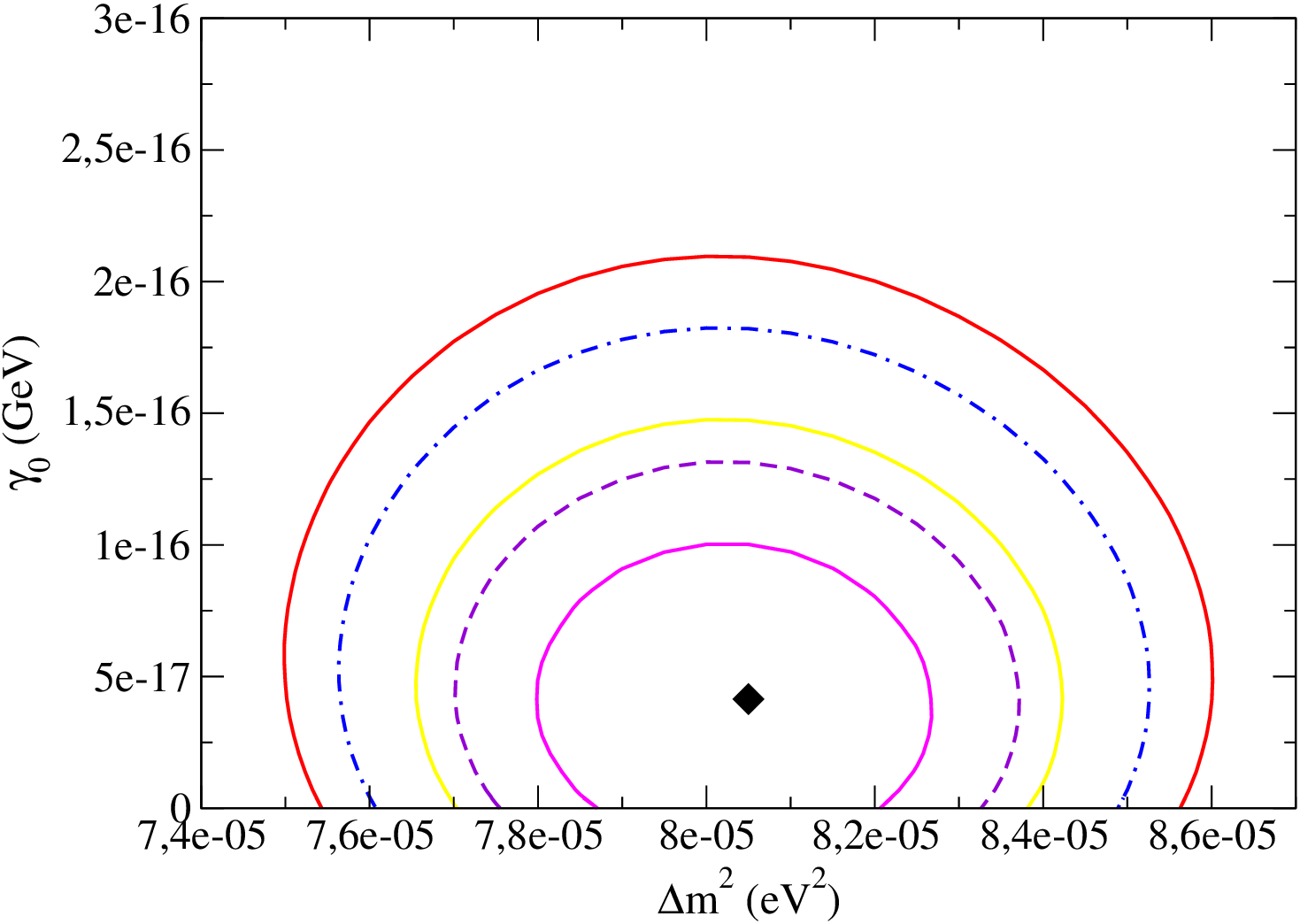}
\caption{Confidence Level curves for $n=1$. The curves correspond to 68.27\%, 90\%, 95\%, 99\% and 99.73\% C.L.}
\end{figure}

\begin{figure}[htb]
\centering
\includegraphics[width=.45\textwidth,]{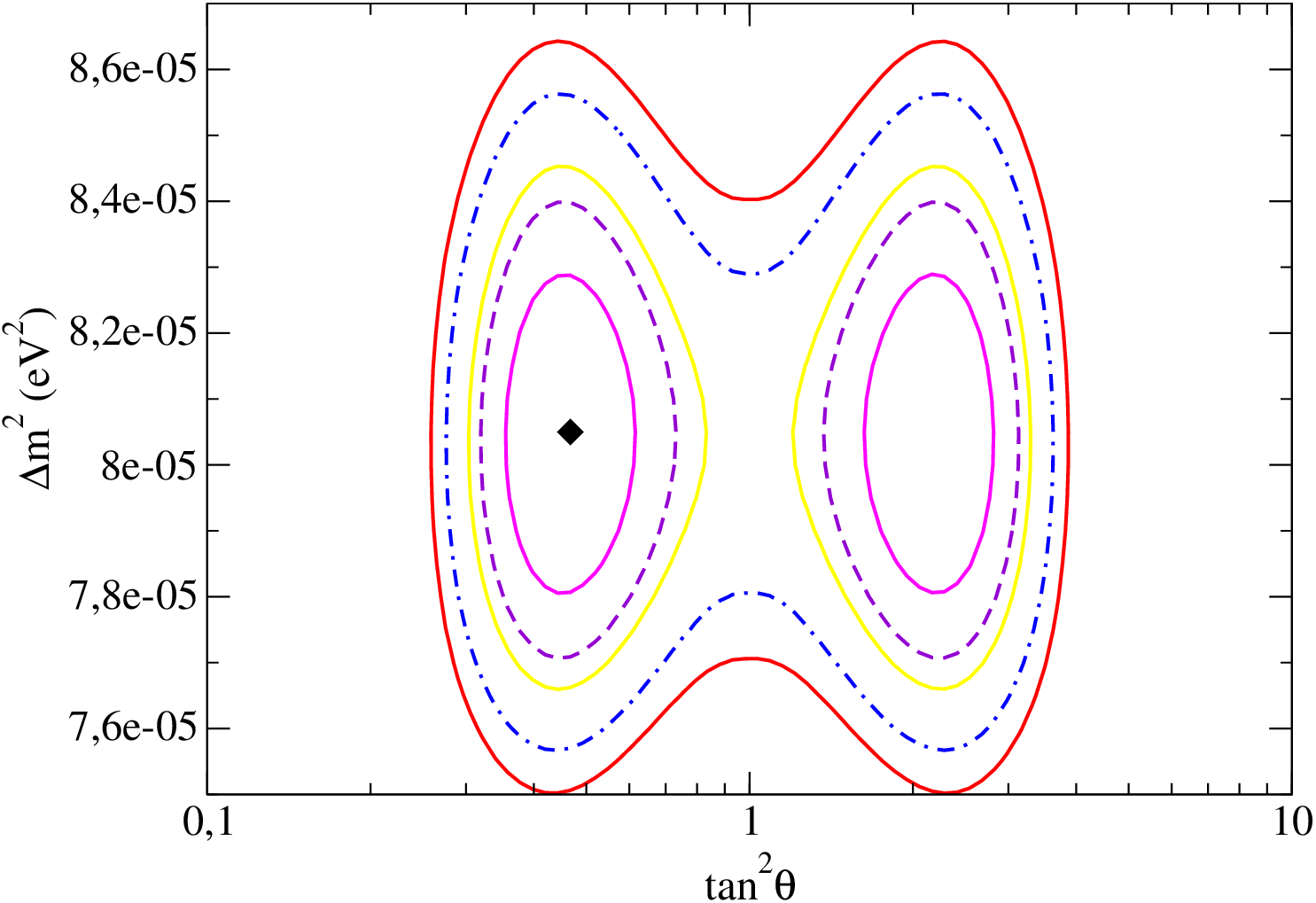}
\caption{Confidence Level curves for $n=-1$. The curves correspond to 68.27\%, 90\%, 95\%, 99\% and 99.73\% C.L.}
\end{figure}

\begin{figure}[htb]
\centering
\includegraphics[width=.45\textwidth,]{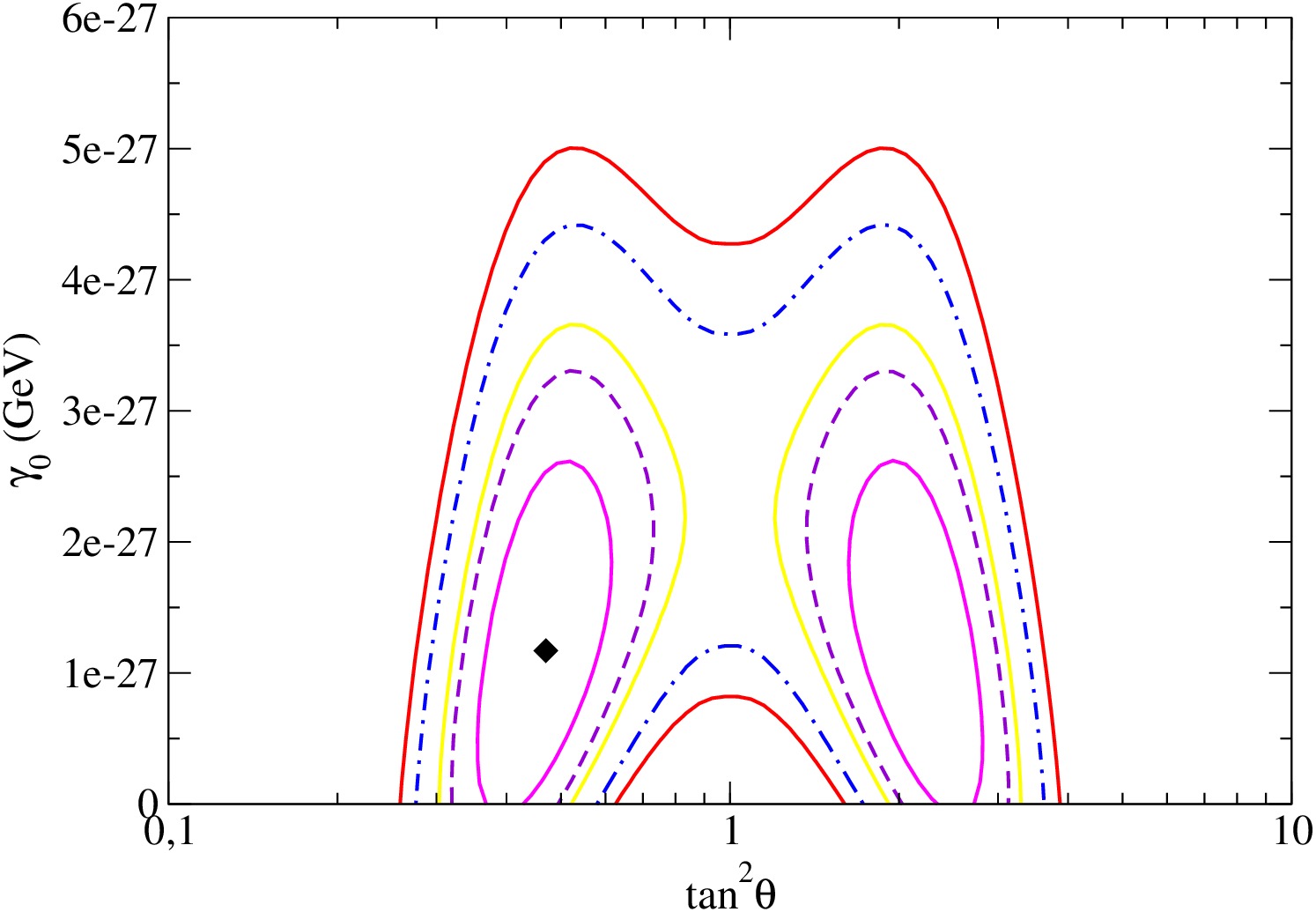}
\caption{Confidence Level curves for $n=-1$. The curves correspond to 68.27\%, 90\%, 95\%, 99\% and 99.73\% C.L.}
\end{figure}

\begin{figure}[htb]
\centering
\includegraphics[width=.45\textwidth,]{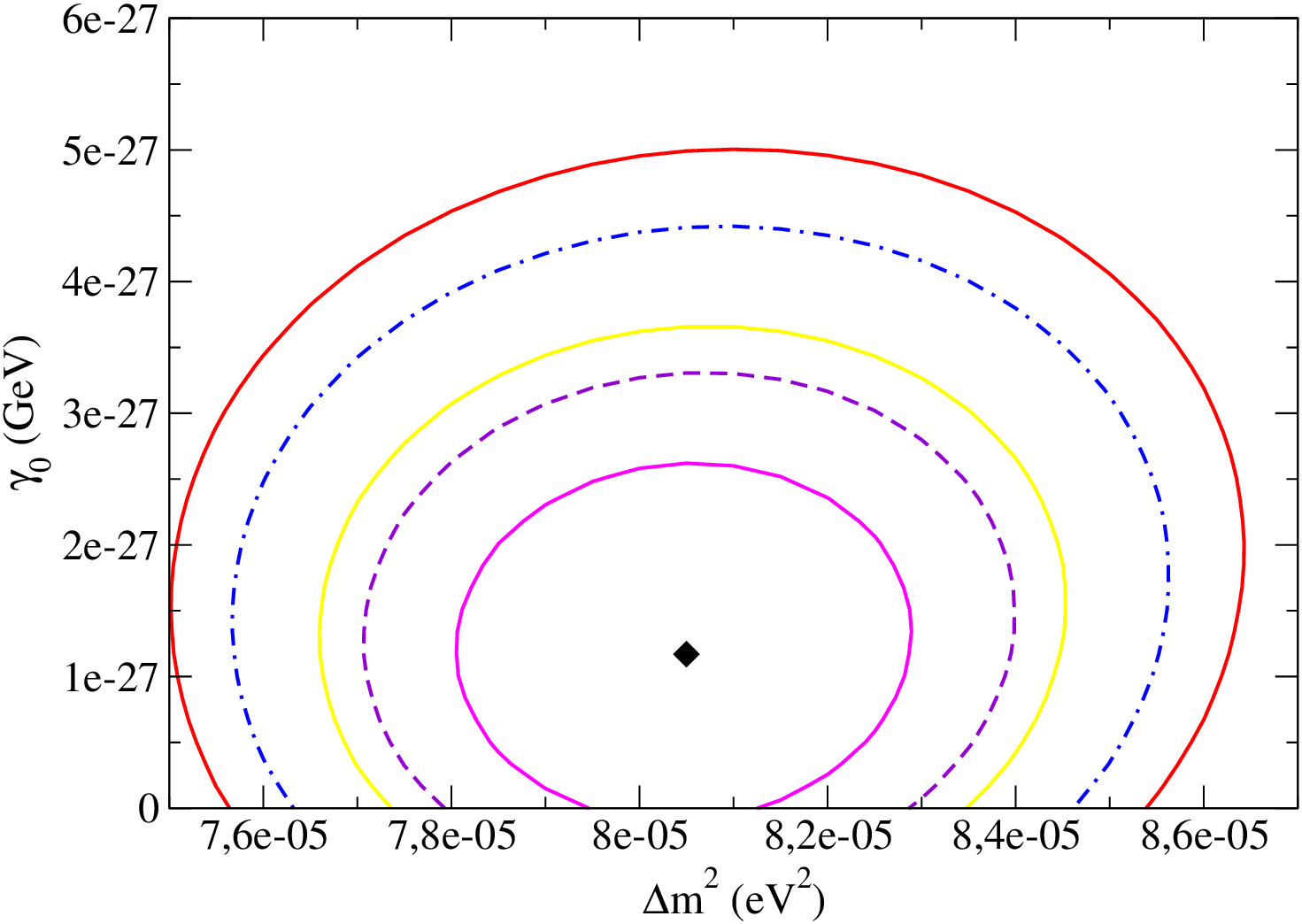}
\caption{Confidence Level curves for $n=-1$. The curves correspond to 68.27\%, 90\%, 95\%, 99\% and 99.73\% C.L.}
\end{figure}

We can see from Figs. 1 to 9 that the decoherence effect does not alter the value of the best fit point for $\Delta m^2$, which is consistent with our previous analysis, since the damping term depending on $\gamma$ acts only on the amplitude of the survival probability.

From the confidence level curves of Figs. 1 to 9, we can obtain limits for the oscillation parameters and for $\gamma_0$, the decoherence parameter. For 95\% C. L. the upper limits on $\gamma_0$ are given in table II:

\begin{table} 

\begin{ruledtabular}
\begin{tabular}{cc}

  $ n=-1$ &  $3.7 \times 10^{-24} GeV$ \\ \hline
 $ n=0$ &   $6.8 \times 10^{-22} GeV$ \\ \hline
  $ n=1 $ &  $1.5 \times 10^{-19} GeV$ \\ 
\end{tabular}
\caption{Upper Limits for $\gamma_0$ in 95\% C.L. with $n=0,1,-1$ }
\vspace{1cm}
\end{ruledtabular}
\end{table}

In order to visualize the effect of the inclusion of decoherence in our study of neutrino oscillations, we can reproduce an important graph originally presented by the KamLAND Collaboration.

Following the same procedure used by KamLAND we used our results to make Fig. 10, which is the result of merging the original graph \cite{klnew} and the graph we made for oscillation with decoherence.

\begin{figure}[htb]
\centering
\includegraphics[width=.45\textwidth,]{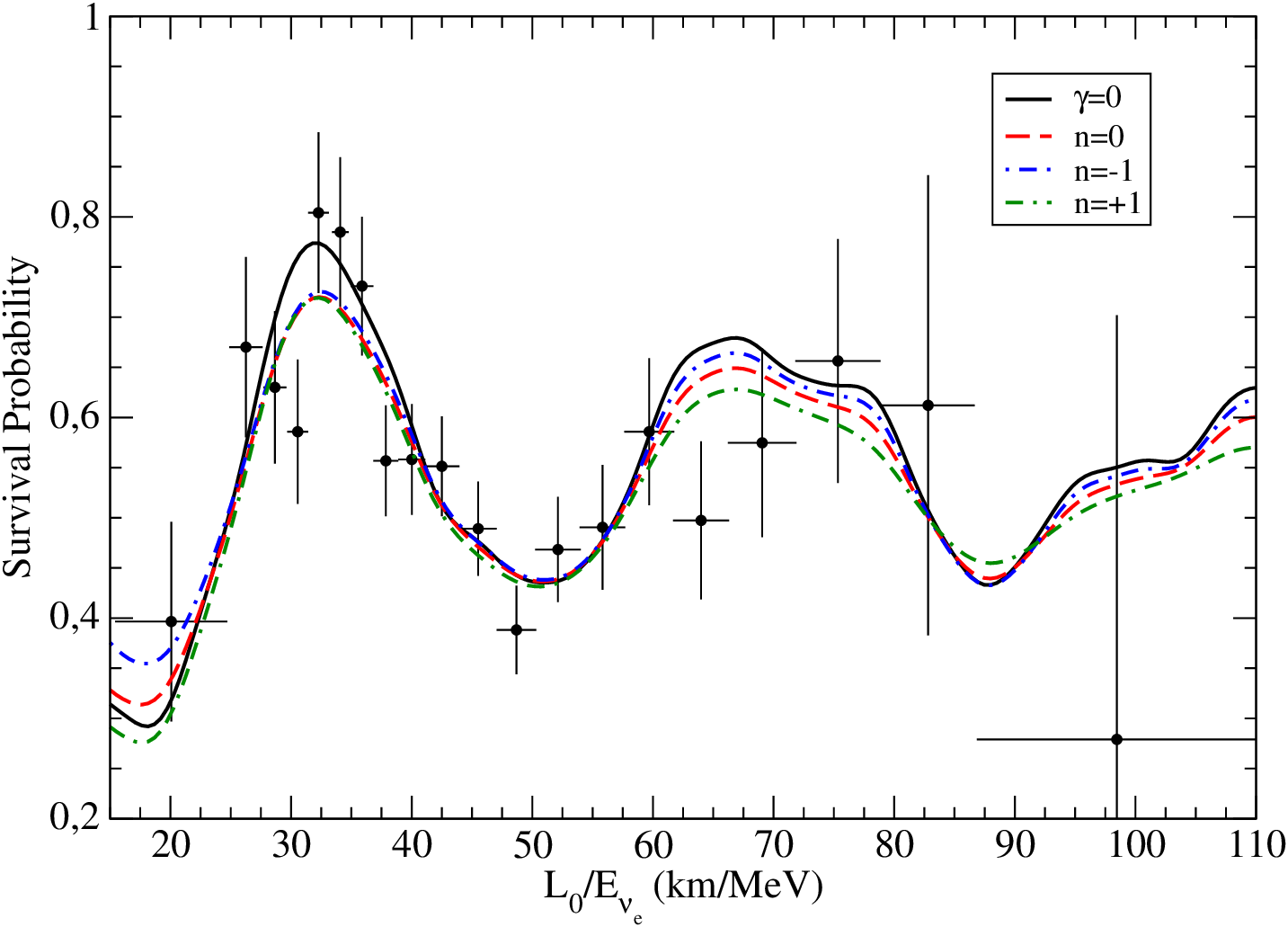}
\caption{Graph made with data from the simulation of our model for oscillation with decoherence  considering best-fit values of the three parameters and the three different values for $n$ in the energy dependence. We also include the KamLAND data \cite{klnew}}
\end{figure}

In Fig. 10 we can see that the fit of the data made from our model of oscillation with decoherence is a good fit of the data, showing a visual confirmation of the analysis provided by the $\chi^2$ Test.

We see that the inclusion of decoherence causes a damping on the oscillation pattern, as we already expected from our theoretical predictions. We can also see that this damping is not too strong for the values of the decoherence parameter best fit points.

\section{Conclusion}

In this work, we treated the appearence of the decoherence effect on neutrino oscillations in a phenomenological approach, studying first Open Quantum Systems in general, and then aplying the results to the case of neutrino oscillation in three families. We analysed the constrains in the model parameters coming from a fit to KamLAND data.

The results were obtained when we considered the most recent set of KamLAND data, provided by Ref. \cite{klnew}, where the number of events were presented in 20 bins. Comparing the value of $\chi^2_{min}$ with the number of degrees of freedom, we saw that including the third parameter, $\gamma$, improves the fit of the data. With $\gamma =0$ we obtained $\chi^2_{min} = 22.96$, and for $\gamma$ as a free parameter (hence 20 experimental points and 3 parameters) we obtained a decrease for $\chi^2_{min}$ of order $\Delta \chi^2\sim 1$. These results are sumarized in table I. We also found a best-fit value with $\gamma \neq 0$.

To support the results of our analysis, giving a more visual way of evaluating the results, we reproduced a graph originally presented by the KamLAND Collaboration, which showed the survival probability \textit{versus} $L_0/E$, which shows clearly the oscillation pattern for the neutrinos. 

Comparing the original graph with our reproduction, which mas made using the best-fit values obtained in our simulation, we saw that our model provided a fit of the data which was indeed in agreement with the experiment uncertainties, as can be seen in Fig. 10.

We also determined new limits for  $\gamma_0$, in  95\% C.L.. The limits are presented in Table II, and were determined based on the confidence level curves made from the most recent set of Kamland data \cite{klnew}.

\begin{acknowledgments}

The authors would like to thank FAPESP, CAPES and CNPq for several financial supports.

\end{acknowledgments}

\end{document}